\documentclass[12pt]{article}
\usepackage{latexsym,amssymb,amsmath}
\usepackage{graphicx}
\begin{document}
%
\thispagestyle{empty}
\title{\bf The inverse crime\\}
\author{
 Armand
Wirgin\thanks{LMA/CNRS, 31 chemin Joseph Aiguier, 13402 Marseille
cedex 20, France, ({\tt wirgin@lma.cnrs-mrs.fr})} }
\date{\today}
\maketitle
\newpage
\noindent {\bf Abstract:} The inverse crime occurs when the same
(or very nearly the same) theoretical ingredients are employed to
synthesize as well as to invert data in an inverse problem. This
act has been qualified as trivial and therefore to be avoided by
Colton and Kress \cite{cokr98}. Their judgement is critically
examined herein.
\newline
\newline
\noindent {\bf Keywords:}  inverse scattering problems; boundary
identification
\newline
\newline
{\bf Abbreviated title:} The inverse crime
\newline
\newline
{\bf Corresponding author:} Armand Wirgin, tel. +33 4 91 16 40 50,
fax +33 4 91 22 08 75, e-mail wirgin@lma.cnrs-mrs.fr
\newpage
\section{Introduction}
Solving inverse problems related to some physical application
(e.g., finding the  constitutive constants of a material from the
reflectivity \cite{wi99}), for a  $N_{p}$-dimensional vector
$\mathbf{p}$   of parameters $\{p_{1},p_{1},...,p_{N_{p}}\}$,
requires matching a theoretical model (parametrized by a vector of
variables of the same nature as the entries of $\mathbf{p}$) of
some observable (reflectivity in the above example) to  $N_{d}$
measurements of the latter embodied in the data vector
$\mathbf{d}=\{d_{1},d_{1},...,d_{N_{d}}\}$.

Often, tests of  the exact or approximate theoretical models
employed in inversion schemes are made with {\it synthetic data}.
Generating the latter also requires a theoretical model, which,
mathematically speaking, can be identical to, or different from,
the one employed in the inversion scheme. In \cite{cokr98}, which
serves as a reference to many workers in the field of inverse
scattering problems, the authors coin the expression "inverse
crime" to denote the act of employing the same model to generate,
as well as to invert, synthetic data. Moreover, they warn against
committing the inverse crime, "in order to avoid trivial
inversion" and go on to state: "it is crucial that the synthetic
data be obtained by a forward solver which has no connection to
the inverse solver".

These assertions raise the following questions: i) what does the
term "no connection" mean? ii) provided a definition can be
provided, what kind of reconstructions of the unknown parameters
can one obtain when there is "no connection" between the forward
and inverse solvers? iii) should the inverse crime always be
avoided? and iv) are inverse crime inversions always trivial?

These questions are difficult to address when $N_{p}$  and/or
$N_{d}$ are larger than one. However, the case $N_{p}$=$N_{d}$=1
is interesting in its own right, and may furnish insight as to
what may occur in inverse problems involving multidimensional
$\mathbf{p}$ and $\mathbf{d}$.
\section{Representations of the forward and inverse solvers
(predictor and estimator)} The inverse problem studied in this
investigation is to determine from $p:=p_{1}\in\mathbb{R}$ from
$d:=d_{1}\in\mathbb{C}$ . What in \cite{cokr98} is called the
"forward solver" is here termed the {\it predictor}, i.e., it
involves predicting the response of a physical system with the
help of some mathematical model incorporating the parameter one
wishes to recover. We denote the predictor by $\Phi(\varphi)$,
wherein $\varphi$ is a variable, of the same nature as $p$, which
can take on the value $p$. What in \cite{cokr98} is called
"inverse solver" is here termed the {\it estimator}, i.e., it
involves estimating the response of some physical system with the
help of the same (as previously) or another mathematical model
incorporating a variable usually having the same physical and
mathematical attributes as the parameter one wishes to recover. We
call this $E(\varepsilon)$, wherein $\varepsilon$  is the variable
analogous to $\varphi$. Thus, for a single experiment, we
synthesize the single piece of data by $d=\Phi(p)$ and seek to
recover $p$ with the help of the estimator $E(\varepsilon)$,
wherein $\varepsilon$ varies over a certain range, hopefully
including $p$.

In fact, we do this for a continuum of experiments involving the
single pieces of data $d=\Phi(p)$, wherein $\varphi$ varies over
the range
\begin{equation}\label{2.1}
  1>\varphi>0~.
\end{equation}
We therefore expect $\varepsilon$  to lie in the same range
\begin{equation}\label{2.2}
  1>\varepsilon>0~,
\end{equation}
and expand the predictor and estimator in the following  series :
\begin{equation}\label{2.3}
  \Phi^{(N)}(\varphi)=\sum_{n=0}^{N}b_{n}f_{n}(\varphi)~~,~~
  E^{(M)}(\varepsilon)=\sum_{m=0}^{M}a_{m}f_{m}(\varepsilon)~,
\end{equation}
(wherein the $f_{m}(\zeta)$  are complex, continuous functions of
the real variable $\zeta$  satisfying $1>\zeta>0$).
 In order to
account for the inverse crime in the easiest manner, it is most
convenient to take
\begin{equation}\label{2.4}
  b_{m}=a_{m}~~;~~m=0,1,2,...~.
\end{equation}
However, we also want to account for the case in which  the
predictor and estimator are {\it different} (i.e., "not connected"
in the language of \cite{cokr98}); this is done by taking $M$  to
be different from $N$.
\section{The comparison equation for the recovery of the parameter
from the data} The usual way to solve the inverse problem is to
minimize some functional of the discrepancy between the estimator
and the predictor. Since, at present, we seek only one parameter,
it is easier to just solve the so-called comparison equation
\begin{equation}\label{3.1}
  K^{(M,N)}(\varepsilon;\varphi):=E^{(M)}(\varepsilon)-\Phi^{(N)}(\varphi)=0~,
\end{equation}
for $\varepsilon$, with the understanding that the reconstruction
will be perfect if $\varepsilon$  turns out to be equal to
$\varphi$.

Note that this way of handling the inversion is equivalent to the
usual way involving minimization, since the zeros of
$K^{(M,N)}(\varepsilon;\varphi)$ are identical to the values of
$\varepsilon$ for which the discrepancy functional attains its
minima.

Note also that, on account of (\ref{2.3}) and (\ref{2.4}),
$\Phi^{(N)}(\varphi)$ can be replaced by $E^{(N)}(\phi)$ in
(\ref{3.1}).
\section{General features of solutions of the comparison equation in
the context of the inverse crime} As stated previously, the
inverse crime corresponds to $M=N$, so that
\begin{equation}\label{4.1}
  K^{(M,M)}(\varepsilon;\varphi):=E^{(M)}(\varepsilon)-\Phi^{(M)}(\varphi)=0~.
\end{equation}
It may be thought that this equation possesses {\it at least one
solution}, i.e., the correct ("trivial inverse" in the language of
\cite{cokr98}) solution
\begin{equation}\label{4.2}
 \varepsilon=\varphi~,
\end{equation}
but this is not necessarily so (inspite of what is stated in
\cite{cokr98}), when, for example, $M=0$  and  $f_{0}$ is a
constant (i.e., does not depend on its argument)).

To go deeper into the features of the inverse crime, we must be
more specific about the functions  $f_{m}$. We make two choices,
one of which is abstract and the other a feature of a real-life
inverse problem.

The first choice is:
\begin{equation}\label{4.3}
f_{m}(\zeta)=\zeta^{m}~,
\end{equation}
by means of which we obtain
\begin{equation}\label{4.4}
  K^{(M,M)}(\varepsilon;\varphi)=
  \sum_{m=0}^{M}a_{m}(\varepsilon^{m}-\varphi^{m})=0~.
\end{equation}
This generally non-linear comparison equation has at least one
solution (i.e., the "trivial inverse"  $\varepsilon=\varphi$) as
long as $M\geq1$. The solution is unique when and only if $M=1$
(we are excluding the absurd choice $a_{1}=0$). Otherwise (i.e.,
$M\geq2$) the polynomial equation (\ref{4.4}) possesses $M$  roots
which are usually not degenerate. For instance, when $M=2$, the
two roots are:
\begin{equation}\label{4.5}
\varepsilon=\varphi,~~,~~\varepsilon=-\varphi-a_{1}/a_{2}~.
\end{equation}
Since the second root is not generally equal to the "trivial
inverse", {\it the inverse crime is not trivial in this case}. Of
course, this argument and conclusion carry over to situations in
which  higher-order models are employed.

Consider next the second choice whereby $M=0$,~$a_{0}\neq 0$, and
\begin{equation}\label{4.6}
f_{0}(\zeta)=\exp(-2ik\zeta)~,
\end{equation}
with $k$  a real constant and $i:=\sqrt{-1}$. This is the exact
model of reflection of  a plane wave $\exp(-ikx_{3})$
normally-incident on a flat mirror  $x_{3}=\varphi~;~\forall
x_{1}\in\mathbb{R}~,~\forall x_{2}\in\mathbb{R}$. For instance, if
the boundary condition is of the Dirichlet type (corresponding to
the case in which the electric field is wholly tangential to the
boundary and the latter covers a perfectly conducting medium),
then $a_{0}=-1$ \cite{wi99}. With this choice, and the observable
being the reflectivity,  the comparison equation (\ref{4.1})
becomes
\begin{equation}\label{4.7}
K^{(0,0)}(\varepsilon;\varphi)=a_{0}[\exp(-2ik\varepsilon)-
\exp(-2ik\varphi)]=0~.
\end{equation}
This is equivalent to the non-linear equation
$\sin[k(\varepsilon-\varphi)]=0$  the solutions of which are:
\begin{equation}\label{4.8}
\varepsilon=\varphi+n\pi/k~;~n\in\mathbb{Z}~.
\end{equation}
Once again, we obtain not only the "trivial inverse"
$\varepsilon=\varphi$, but a number (infinite) of other solutions.
The latter betray the fact that, even in the context of the
inverse crime, the inverse problem of locating the height
$\varphi$ of the mirror from one measurement of the reflectivity
is  an ill-posed problem (due to the non-uniqueness of the
solutions \cite{ha23},\cite{bu93}). This finding is at odds with
what is written in \cite{cokr98} concerning a boundary recovery
problem solved by committing  the inverse crime: "Hence, it is no
surprise...that the surface $\partial D$ is recovered pretty well"
(here  $\partial D$ is the boundary of the mirror, or in
particular its location), since, in our example, not only do we
recover the location of the mirror, but also the locations of
other planes on which the boundary condition is satisfied (this
being a surprise).

The latter example illustrates another feature of the inverse
crime: its {\it usefulness}.  To be more specific, (\ref{4.8})
tells us that we can actually obtain the location of the mirror
unambiguously from {\it two or more} experiments conducted for two
or more values of $k$ (i.e., for $\geq 2$  values of the frequency
of the incident wave), since the only value of $\varepsilon$ in
(\ref{4.8}) that is independent of the frequency is the correct
value $\varepsilon=\varphi$. This idea has been put to use
recently in a  more complicated shape identification problem
\cite{ogsc01}.
\section{The "no connection" issue}
As stated previously, in the framework of this paper, an estimator
that is different from the predictor means that $M\neq N$  so that
the comparison equation is:
\begin{equation}\label{5.1}
K^{(M,N)}(\varepsilon;\varphi):=
\sum_{m=0}^{M}a_{m}f_{m}(\varepsilon)-
\sum_{m=0}^{N}a_{m}f_{m}(\varphi)=0~.
\end{equation}
The term  "no connection" can be understood to mean that $M$  be
very different from $N$, but this may not be necessary for certain
ranges of the estimator and predictor.

To begin the discussion, consider again the case $M=0$  such that
$f_{0}$ is a constant (i.e., does not depend on its argument).
Then $\varepsilon$ does not appear in (\ref{5.1}) which means the
non-existence of a solution.

Lest this type of estimator appear to be too extreme, consider the
choice of $f_{m}$  given in (\ref{4.3}), and the  case in which
the estimator is linear and the predictor quadratic. Then
(\ref{5.1}) yields
\begin{equation}\label{5.2}
\varepsilon=\varphi+(a_{2}/a_{1})\varphi^{2}~.
\end{equation}
What is remarkable about this result is the {\it uniqueness} of
the solution (due to the linearity of the estimator). However, the
relative error of the reconstruction
$\delta=|(\varepsilon-\varphi)/\varphi|$   (=0 for the "trivial
inverse") is
\begin{equation}\label{5.3}
\delta=|(a_{2}/a_{1})\varphi|~,
\end{equation}
and this is small only if $|(a_{2}/a_{1})|$ and/or $|\varphi|$
are/is small. This shows that it may be impossible to obtain a
solution, whose relative error lies below some prescribed
threshold, when employing an estimator (linear in this example)
that has "no connection" to the predictor (quadratic in this
example). A similar finding can, of course, be obtained for other
unconnected estimator/predictor pairs. In other words, if one
wants to recover the unknown parameter very accurately he
shouldn't follow the recommendation \cite{cokr98} of choosing an
estimator that has "no connection" with the predictor. In general,
the larger is the functional difference between the estimator and
the predictor, the larger is the relative error of the inversion
\cite{scwi96}.
\section{Discussion}
Let us return to the four questions raised in section 1.
Concerning the meaning of the term ''no connection'', we proposed
that this be materialized by a difference in the number of terms
in a power series representation of the predictor and estimator.
Of course, we could have proposed more radical differences, but it
is hardly conceivable that they would yield a  result that is
better than the one ((\ref{5.3})) of section 5.

The response to the  question concerning the kind of
reconstructions one can obtain when there is "no connection"
between the predictor and estimator is provided eloquently by
(\ref{5.3}), i.e., in such a case one can obtain reconstructions
of $\mathbf{p}$ that have "no connection" to the actual
$\mathbf{p}$.

The reponse to the question as to whether the inverse crime should
always be avoided is, negative, since committing this crime can,
at the very least, reveal the non-uniqueness of the inverse
problem. It could be argued that this non-uniqueness (certainly a
negative feature) is somehow induced by the inverse crime, with
the implication that abolishing this crime would make the solution
unique. However, non-uniqueness is neither a necessary, nor
specific,  feature of the inverse crime as is illustrated by the
fact that by taking a linear estimator and linear  predictor one
obtains one and only one solution, the so-called "trivial
solution" $\varepsilon=\varphi$.

A last comment on this question: it is rather strange to employ
the disparaging term "trivial" to qualify what is, after all, the
very objective  of solving an inverse problem, i.e.,
reconstructing as accurately as possible the unknown parameters
(in our case, obtaining $\varepsilon$   as close as possible to
$\varphi$).

The response to the question as to whether inverse crime solutions
are {\it always trivial}, is  provided by (\ref{4.5}) for the
abstract example, and by (\ref{4.8}) for the concrete example,
i.e., not only does committing the inverse crime lead to the
"trivial solution" $\varepsilon=\varphi$ , but also to one or more
other solutions whose existence might be observed in a full-blown
numerical procedure, but not necessarily revealed by a
non-inverse-crime analysis.

The material provided herein, based essentially on three examples,
does not pretend to be universal. Obtaining one parameter from one
piece of data does not resume the way things are usually done in
the field of inverse problem solving. In fact, often one relies on
more data than the number of unknown parameters, but even this
does not always resolve the non-uniqueness issue \cite{de78}. This
is particularly so in an abstract setting (as for the first
example in section 4), but, as we have shown, when there are
physical reasons why a solution should be invariant to the change
in some physical parameter (e.g., the location of a mirror should
be invariant to the frequency of the probe radiation) then two
measurements (at different frequencies) are certainly better than
one measurement to distinguish the correct solution from false
solutions, provided, of course that the error (experimental, or
with respect to the estimator when employing synthetic data) of
the two (or more) pieces of data are of the same order.

When, as is usually the case, one wishes to recover more than one
parameter from more than one pieces of data, then a mathematical
analysis of the type proposed herein rapidly becomes very
complicated, but would, if it were carried out, certainly provide
a useful contribution to the field, although it can hardly be
expected that the last sentence of section 5 would be contradicted
in situations more complicated than the ones analyzed herein.

A final question, not raised in section 1, is whether there is a
risk (if such is one's concern) of committing the inverse crime in
practical problems, this meaning problems appealing to {\it real}
(as opposed to synthetic) data. We see no such risk since {\it the
predictor for real data is unknown}. The {\it worst} (if one
apprehends punishment for committing a crime, but the {\it best}
otherwise) that can happen  is that the experiment be done very
properly and the model employed in the estimator very nearly mimic
the physics of the experiment, in which case the chances are great
that one will accurately recover the sought-for parameters of the
object under study. But, as we have shown herein, one might also
recover artifacts \cite{ogwi01} that will have to be eliminated by
some sort of strategy such as performing other well-chosen
experiments (this choice could be facilitated by an analysis
relying on data synthesized by a predictor that is functionally
equivalent to the estimator, i.e., by committing the inverse
crime)  or incorporating a priori information about the object at
the comparison stage (this is usually what is done in the
procedure called regularization \cite{tiar74}).

\end{document}